# Model-independent Evidence for an Increase in the Mean Mass of Cosmic Rays above 3 EeV

A A Watson[1]

[1] School of Physics and Astronomy, University of Leeds, Leeds, UK

**Abstract.** Measurements of the Elongation Rate of the depth of shower maximum above 1 EeV are reviewed. There is evidence, from four independent estimates of this rate, made in the two hemispheres using three different techniques, for a decrease in the Elongation Rate above ~3 EeV, as first discovered by the Pierre Auger Collaboration over 15 year ago. Unless there is a dramatic change in the hadronic physics above this energy, the mean mass of the primary cosmic rays must increase as a function of energy, well into the decade beyond 10 EeV. To estimate the mass, the use of hadronic models is required, the accuracy of which remains uncertain. However, the possibility of a dramatic change in the hadronic physics appears unlikely, and would be inconsistent with data from the Auger Collaboration on the mass composition in the range 3 to 10 EeV, and on the anisotropy of arrival directions above 8 EeV. Both of these conclusions are insensitive to uncertainties in the shower models. Some remarks are made about the belief, still firmly held by some, that the highest-energy cosmic rays are dominantly protons.

## 1 Introduction

The talk, on which this paper is based, was prepared in the hope that it would re-activate discussion of the topic of the Elongation Rate at high energies within the Auger/Telescope Array (TA) Working Group on Mass Composition. In fact, during UHECR2022, Bergman [1] presented evidence showing that measurements of the depth of shower maximum by the TA group are now considered as being consistent with the mixed mass observed by the Auger Collaboration. Thus, significant progress has been made, and, although the conclusions are restricted to energies below ~10 EeV, and apply only for some hadronic models, a major step forward appears to have been made: the joint publication promised by the Working Group is therefore eagerly awaited.

Some of what follows is a replay of arguments set out in [2] in which the claim made by Sokolsky and D'Avignon [3] that the Elongation Rates measured in the Northern and Southern Hemispheres are different was refuted. Additionally, the likelihood of a change in the hadronic physics, and the evidence against it are briefly discussed, and the long-standing belief of some that protons dominate at the highest energies is examined.

The TA Collaboration has not yet made comparisons of their $X_{max}$ results with model predictions beyond 10 EeV as they hold the view that ~300 events above 10 EeV are too few to make reliable deductions [4]. It was argued in [2] that *only* TA data from the Black Rock and Long Ridge detectors should be used in such studies, and this is what has now been adopted in [1, 4]. In this TA data set, there are 3330 events above 1 EeV: the measurements [4] are shown in figure 1, alongside those from a recent report from the Auger Collaboration [5].

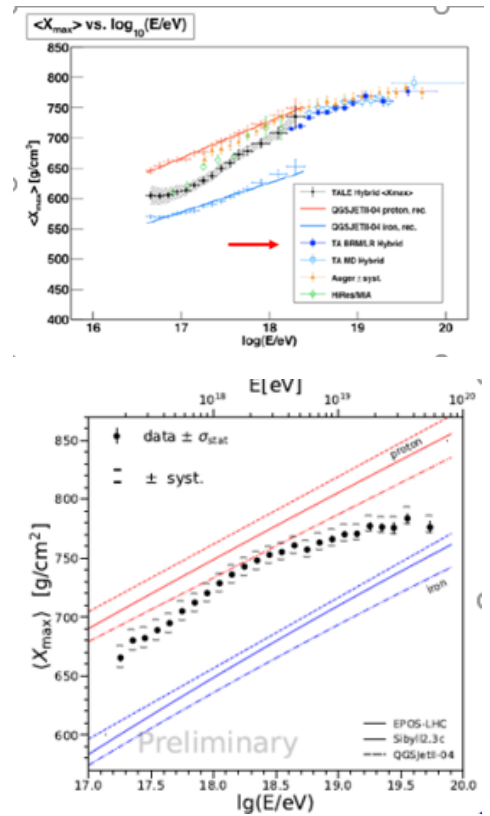

**Fig.1**. Recent measurements of the depth of shower maximum from the Telescope Array Collaboration (upper plot) [4], and from the Auger Collaboration (lower plot figure) [5]. Note that different models have been used by the Collaborations for the comparisons of data with models. The QGSJet II-04 model gives a poor description of some features of data from the Auger Observatory. For the Auger data, the Elongation Rate above 3 EeV is $26 \pm 2.5$ g cm$^{-2}$/decade.

Conclusions about primary mass drawn from plots such as those of figure 1 are dependent on the hadronic model assumed. This is a long-standing, and continuing, problem, as illustrated in figure 2 where data from the Fly's Eye detector are compared with the predictions from different models. The upper-plot is what was shown in [6], nearly 30 years ago, when the QCD Pomeron model was 'state-of-the-art'. This model was used to draw the conclusion that a light component, usually assumed to be protons, was becoming dominant at the highest energies.

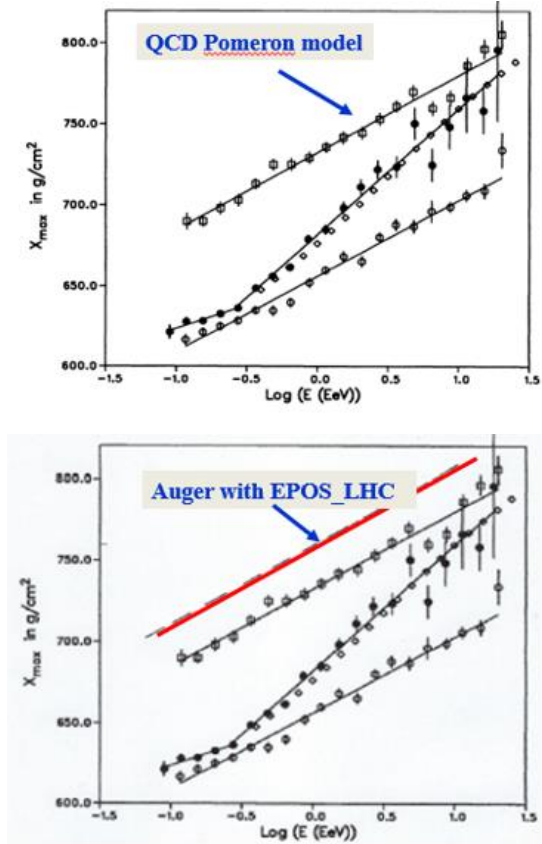

**Fig.2.** Data from [6] are shown in both plots. The different conclusions that can be drawn according to which model is adopted are evident (see discussion in text).

This plot has, I believe, had profound influence on the thinking of many people, particularly phenomenologists. In the lower plot, the same data are compared with the predictions of a current hadronic model (EPOS-LHC) from which larger values $X_{max}$ are found. These predictions shown were made with the selection procedures used to derive values of $X_{max}$ from the data of the Auger Observatory, and would differ by a few g cm$^{-2}$ from what is shown were the Fly's Eye selection procedures adopted instead. However, the important point is that the conclusion would be the same: the mass derived is strongly dependent on the hadronic model selected, and, above ~0.1 EeV, deductions from LHC data require extrapolation.

In [2], it was pointed out that the break in the Elongation Rate deduced from the *ensemble* of data amassed from projects carried out in Utah over the years [3], had been discussed previously [7]. The break is also seen by the Yakutsk group, using measurements of the lateral distribution air-Cherenkov radiation, and in measurements using the risetimes of the signals in the water-Cherenkov tanks of the Auger Observatory. These three results are summarised in figure 3. A detailed discussion of these plots can be found in [2].

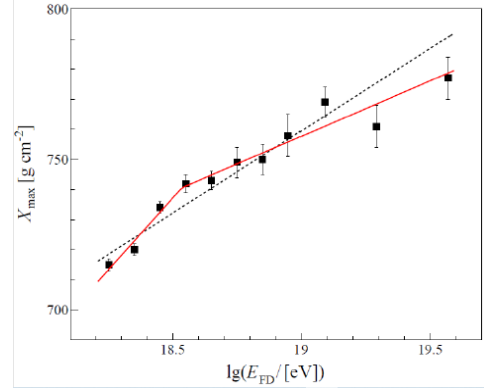

**Fig.3a.** $X_{max}$ vs Energy from the Black Rock/Long Ridge Observations by the Telescope Array Collaboration. The dashed line is a fit to the points with a single straight line (Elongation Rate = 54 ± 3 g cm$^{-2}$/decade). The full line is a fit with a break and the best-fit straight lines found simultaneously. The Elongation Rate above the break at 3 EeV is 37 ± 5 g cm$^{-2}$/decade. See [2] for further discussion.

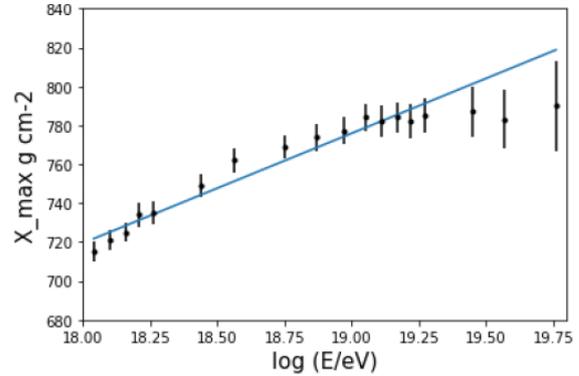

**Fig.3b.** Data of $X_{max}$ vs log E/eV from Yakutsk. The Elongation Rate above 3 EeV is 28 ± 7 g cm$^{-2}$/decade. See [2] for details.

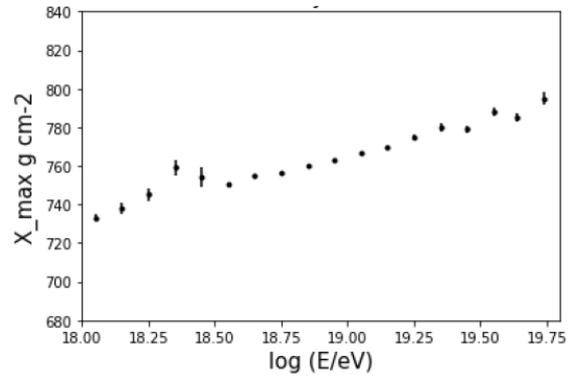

**Fig.3c.** Estimates of $X_{max}$ as a function of energy made at the Auger Observatory using the risetime technique. The Elongation Rate above 3 EeV is 33.2 ± 0.7 g cm$^{-2}$/decade. Further details are given in [2].

The weighted averages of the Elongation Rates (ER2) above 3 EeV for the two hemispheres are:

Northern hemisphere:

TA and Yakutsk: ER2 = 34 ± 4 g cm$^{-2}$/decade

Southern hemisphere:

Two independent measurements from the Auger Observatory: ER2 = 32.7 ± 0.7 g cm$^{-2}$/decade.

It is evident that the claim advanced in [3] that the Elongation Rates are different in the two hemispheres is not supported.

It also follows that the mean mass of the primary particles increases as the cosmic-ray energy rises, unless there is a dramatic change in features of the hadronic interactions above the centre-of-mass energies reached at the LHC (equivalent to ~0.1 EeV for p-p collisions in cosmic rays). Exactly what mean mass is deduced depends strongly on the choice of models (see figures 1 and 2). While the conclusion drawn here is certainly not original (see for example [8]), it is worth restating, as it does not depend on assumptions about hadronic physics. Also, there remain many who do not accept this conclusion.

The situation with regard to the best model to adopt remains fluid. For example, at this meeting, we learned from Pierog [9] that it may be necessary to include the production of quark-gluon plasmas in the calculations. Of course, a dramatic rise in the p-p cross-section and multiplicity of pions resulting from collisions can also be invoked to enable a proton to mimic a heavy nucleus.

## 2 The possibility of changes in the hadronic physics

The idea that features observed in cosmic-ray data have their origins in changes in hadronic physics, rather than providing evidence of an astrophysical feature, has a long history. In 1958, Kulikov and Khristiansen [10] reported a break in the number spectrum of extensive air showers, now known as the knee. They proposed that this was evidence for an astrophysical feature, specifically a demonstration in favour of an extragalactic origin of cosmic rays above $10^{16}$ eV. However, interpretations of this measurement, and of later confirmatory ones, in terms of a change in hadronic physics persisted for over 40 years (e.g. [11]). It was not until 2002, that the KASCADE group ruled out this possibility, *without resorting to models,* by demonstrating that the number spectrum of muons steepened around the knee, but only in showers that were rich in electrons (i.e. shower produced by light nuclei) [12]. A similar approach was later used by the KASCADE-Grande Collaboration to show that there is a second knee at around 0.1 EeV, associated with the loss of heavy nuclei from the primary beam [13]. I find such *model-independent conclusions* very compelling: trends are demonstrated that can be interpreted rather easily with an elementary understanding of the principles of shower development.

Also, one should be alert to the over-confidence of theorists. Up until 1973, major authorities, including Heisenberg, asserted that the proton-proton cross-section would remain constant with energy beyond a few GeV. This was a strongly held view, based on Regge Pole theory, that the p-p cross-section had reached a finite asymptotic value. Of course, a study at the CERN ISR proved this theoretical prediction to have been incorrect (see figure 4, taken from [14]).

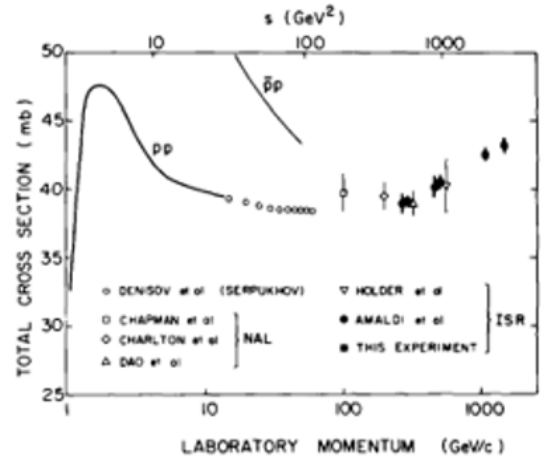

**Fig.4**. Rise in the total p-p cross-section as reported in [14].

The unexpected nature of this result has been discussed by Amaldi [15], while in [16] Matthiae, who was heavily involved in developing the 'Roman pots' used for these measurements, describes the rise in the cross-section as *'a startling discovery'*. There had been earlier claims by Grigorov et al. [17] and by Yodh et al. [18] of a rise in the nucleon-air cross-section deduced by comparing the rate of unaccompanied nucleons at high altitude with the flux of protons at the top of the atmosphere. However, it was difficult to assess the systematic uncertainties in these pioneering cosmic-ray experiments, particularly because of limited knowledge of the primary spectrum of protons. Indications of a rise in the p-air cross-section were certainly present, with the results in reasonable accord with the accelerator results.

Whether such surprises await us above 1 EeV is, of course, unknowable. At a recent conference [19], Unger reported on work carried out by Engel, Ulrich and himself, in which it was argued that the change in the p-air cross-section by the factor of two required to explain the change in the Elongation Rate, if protons persist, is very extreme. The minimum change required for the p-p cross-section to obtain such a large value for the p-air cross-section with a Glauber model of nucleon-nucleus scattering is about a factor of three (at a centre-of-mass energy within about an order of magnitude of that reached at the LHC). Romanopoulos, Pavliou and Tomaras [20] have recently discussed the same problem.

There is also experimental evidence against such a change. Firstly, in the region of the ankle, 3 to 10 EeV, the Auger Collaboration [21] has presented an analysis showing that the correlation between the depth of shower maximum and the signal in the water-Cherenkov detectors supports a mixed composition with the atomic mass greater than 4, thus disfavouring claims of a pure composition in this energy range. This conclusion does not depend on different models of the hadronic physics.

Secondly, it may well be that evidence from anisotropy studies, along with better a understanding of the magnetic fields lying between the sources and the earth, will make it possible to rule out such a possibility in a rather direct manner. I see this line of attack as being as fruitful as that used in [12], which ruled out explanations of the knee in terms of changes in hadronic physics. At present, features such as the dipole measured in the distribution of arrival directions by the Auger Collaboration above 8 EeV, with very high significance [22], are found to incompatible with a pure proton composition [23]. These experimental results [21 and 22] should be taken as a clear refutation of a pure proton composition.

## 3 The Belief in Protons at the Highest Energies

The idea that the highest-energy cosmic rays are dominantly protons has enormously influenced thinking about the origin of these particles for over 60 years. As long ago as 1955, the distinguished cosmic ray and particle physicist, G Cocconi wrote [24]:

*"We remain with the dilemma: protons versus heavy nuclei. A clear cut decision cannot be reached yet. I believe that up to the highest energies the protons are the most abundant in the primary cosmic rays. …………………… However, I must confess that a leak proof test of the protonic nature of the primaries at the highest energies does not exist. This is a very important problem. Experimentally it is quite a difficult problem."*

The final sentence has proved to be a prophetic understatement. Also, it is worth keeping in mind that, in 1955, the detection of particles with energies above 0.1 EeV had barely been achieved.

At that time, estimates of the primary energy were made using a calorimetric approach, assessing the energies carried by the electrons and muons separately, and by including ionisation losses, which dominate, in both the atmosphere and below ground level. Estimates based on models of particle interactions were just being developed. Particles above 0.1 EeV were soon to be observed, principally through the efforts of the MIT group [25, 26]. The view of Linsley on the primary mass, based on observations with the small muon detector included in 8 km$^2$ array at Volcano Ranch, was that the muon distribution above ~0.1 EeV implied that either protons or nuclei of the Fe group dominate, though he also had arguments that appeared to favour protons [26]. The Volcano Ranch results were for the range 0.1 EeV to 1 EeV.

That the proton idea became so embedded in the thinking of the cosmic ray community is rather alarming. There has been an almost Orwellian group-think, possibly largely stimulated by the results from the Fly's Eye experiment [6] shown in figure 2. That result, dependent as it was on assumptions on the choice of hadronic model, appears to have carried more weight than other indications from astrophysical ideas about acceleration.

Cocconi's view was based on a number of observations and the theories that were available at a time when there was even deep suspicion, from many low-energy cosmic-ray physicists, as to the reality of the shower phenomenon. The idea that a shower might be induced by several lower energy photons arriving simultaneously was seriously considered. Cocconi [24] used the lack of lateral spread of the nucleonic component in showers as an argument against heavy primaries. Also, Greisen's view [27], that if Fermi's theory for an acceleration mechanism was correct, then lighter nuclei would be favoured, as their cross-section for collision was smaller, would almost surely have been well-known to Cocconi who was working along side Greisen at Cornell.

The idea that the higher energy particles might be of extragalactic origin, first discussed in [10], led Gerasimova and Rozenthal [28] to point out that nuclei with energies of 10$^{16}$ eV/nucleon would be subject to photodisintegration in intergalactic space, and, with the photon fields assumed by them, to the claim that most heavy nuclei would disappear. However, Ginzburg and Syrovatskii argued that the photon density adopted was a least an order of magnitude lower than assumed in [28] so that heavy nuclei would survive [29]. After the theory of diffusive shock acceleration became widely accepted, Hillas [30], in the famous 'Hillas plot', pointed out that the highest energy attained was proportional to Z, the atomic number of the nucleus of interest. Rigidity is a key parameter.

Evidence for the feature in the elongation rate discussed in section 1 began to emerge in data obtained by the Auger Collaboration even before construction of the instrument was completed. In 2007, on the basis of ~1000 events with E > 2 EeV, it was pointed that there was evidence for the mean mass becoming heavier as the energy grew [31]. This indication was reinforced in 2010 when results from approximately twice as many events were reported [32].

That these results were in contradiction with conclusions drawn from the successive projects in Utah (Fly's Eye, HiRes - and later Telescope Array) has clearly been part of the problem, and certainly key people in the field, such as V S Berezinsky and E Waxman, were extraordinarily reluctant to consider other than pure protons at the highest energies. Around the time of the publication of [8], in which results based on over 5000 events above 2 EeV are presented, I was invited to a meeting of acceleration specialists at Dunsink. Ireland. I had been charged to set out the experimental situation and I described in critical detail the evidence available. I recall being told by M Lemoine that he preferred to try to find ways of accelerating protons to 100 EeV rather than iron nuclei as it was a more challenging problem. True of course, but not a way to help our understanding, particularly as theorists generally have much more influence than do the gatherers of the data.

Of course, the belief in a dominance of protons at the highest energies also leads to more encouraging fluxes for those planning the detection of UHE neutrinos.

## 4 Take Home Messages

1. There is no evidence for any difference in the Elongation Rate as observed in the two hemispheres.

2. The break in the Elongation Rate above 3 EeV povides *model-free* evidence that the mean mass of the primary particles becomes heavier as the energy increases.

3. Claims that protons are in fact dominant at high energies because of the increase in the p-air cross-section and the multiplicity of pions from such collisions are in disagreement with experimental evidence on the arrival directions of cosmic rays above 8 EeV and on the model-independent evidence for A>4 in the ankle region.

4. The belief that protons dominate at the highest energies should be abandoned.

I am grateful to Michael Unger for clarifying information about the rise in the p-p cross-section at the highest energies.